\documentclass[aps,preprint,groupedaddress,showpacs,showkeys]{revtex4-1}

\usepackage{graphicx}
\usepackage{bm}

\begin{document}


\title{A quasi-conserved particle Monte Carlo model of surface evolution with semi-empirical sputter yield modulated \\ erosion: 1 keV Ar$^+$ sputtering of Si }


\author{Emmanuel Oyewande}
\email[e-mail: ]{eoyewande@gmail.com, oe.oyewande@ui.edu.ng}
\affiliation{Department of Physics,
  University of Ibadan, Ibadan, Nigeria.}


\date{\today}

\begin{abstract}
We introduce a new Monte Carlo model based on a semi-empirical sputter yield parameter in ion-solid energetic collisions. This model circumvents the complexity of the existing statistical, classical and continuum models, most of which are difficult to relate to real experimental parameters, by its semi-empirical nature of direct reliance on the experimental values of the sputter yield. Constrained by this crucial experimental factor, the model then addresses the multidimensional nature of other accompanying physical processes stochastically; thus reducing the complexity of their computation. This model exhibits the experimentally observed features of solid surfaces that evolve under continuous particle irradiation and allows for a way to study the effect of the different mechanisms of the surface morphology and the nature of their interplay in the dynamics of the surface evolution. Our study of the average surface height reveals that it is constant when eroded particles are redeposited but varies linearly with simulation time, when eroded particles are not redeposited. Our studies also show that the roughening process is not significantly affected by re-deposition of eroded material.
\end{abstract}


\maketitle

\section{\label{intro} INTRODUCTION}

Ion-beam surface sputtering (IBSS) is a versatile technique for producing nanostructures on material surfaces \cite{1, 2, 3}, for high-tech opto-magneto-electronic applications. These nano-structures are formed by self-organisation of surface atoms, as the surface undergoes continuous bombardment by energetic particles, into highly ordered nano-ripples, nano-dots, and nano-holes. Their type, size, shape, separation, symmetry/crystallinity, and orientation can be tuned by varying the sputtering conditions such as irradiation time, ion energy, dose, angle of incidence, and substrate temperature.  

Up till now, the level of control attained in the use of this technique for the fabrication of nanostructures defies a fundamental 		understanding of the phenomenon \cite{4}, though its core mechanisms have been identified, which greatly limits a large-scale application of the technique. This knowledge gap is further compounded by recent ascertainment of a decisive influence of contaminants in the self-organisation process. For instance, metal (e.g. Fe and Mo) atoms may be accidentally sputtered off the ion source, vacuum walls and target clamps, made from such metallic materials, and incorporated in the evolving surface \cite{4}, thus playing an essential role in the surface morphology. On the other hand, some recent experiments reported the production of pure nanostructures by IBSS \cite{5}, the morphology of which was uninfluenced by contaminants. 

Thus, it remains unclear whether there is another mechanism which acts differently in the presence or absence of 			          contaminants such that nanostructures are produced anyway, and whether this mechanism plays any role in tuning the characteristics of the nanostructures. Until these issues are resolved in both theory and experiments, the potential of IBSS as a nano-patterning and nanostructure production technique will remain largely untapped. And, the potential of the nanostructures evolved from the process for super/smart-tech applications that would exploit their properties and tunable characteristics will remain largely unutilised. 

Therefore, the importance of theoretical modeling for investigating the mechanisms involved in the IBSS processes as well as their dynamics and interplays cannot be over-emphasised in the quest to achieving a precise control over the above-mentioned properties of the nanostructures produced by IBSS. And the current knowledge gap, despite advances in the theory, motivates a closer scrutiny of existing theoretical models and the development of more sensitive ones. The theory has been widely studied in three major theoretical frameworks \cite{2}, namely: molecular dynamics simulations in the classical physics framework \cite{6, 7}, Monte Carlo simulations in the statistical physics framework \cite{8, 9, 10, 11, 12}, and simulations with continuum models in a mathematical framework of linear and nonlinear partial differential equations \cite{13, 14, 15, 16, 17}. 

However, this is an area where the theoretical framework of statistical physics is particularly suited to reveal interesting and surprising correspondence with other non-equilibrium phenomena. A characteristic feature of this framework is the concept of universality, by which seemingly different and diverse phenomena are governed by the same simple statistical mechanical law because oftentimes the apparent complications of the relevant processes mask well-ordered and far less complicated mechanisms which are inconspicuously present and are ultimately responsible for the law (see Refs. \cite{3} and \cite{18} for reviews, and Ref. \cite{19} for a similar feature in quantum field theory). 

Hence, in this article and with subsequent investigations of universality in view, we introduce a discrete model in the framework of statistical physics, for the simulation of IBSS, taking cognizance of the core mechanisms in IBSS and their influence on the dynamics and self-organisation of surface constituent particles. In this instance, we focus on amorphous surfaces or surfaces amorphised by the ion impacts; a characteristic of semiconductor materials, and a paradigm of which is Si. This model differs from the other statistical models \cite{20, 21, 22, 23} in its semi-empirical nature and material particle conservation.

The rest of the article is organised as follows. We provide background material in section \ref{sec:ExptTheor}, where we review the experimental background of IBSS in section \ref{sec:RevExpt}. In section \ref{sec:RevTheor}, we review a few MD, MC, and MD-MC hybrid models in the literature in order to provide a background to the existing models. The experiments and models reviewed in this paper are just a few of the common ones and the more recent ones the author has come across, and not exhaustive. In section \ref{sec:Probs}, we discuss some open questions and issues on the existing models. In section \ref{sec:Model}, we present our model and discuss its relation to experimental parameters. In section \ref{sec:Results}, we present and discuss our results, and finally, we round up the paper with a conclusion and suggestions for further work.

\section{\label{sec:ExptTheor} Experimental and Theoretical Background}
\subsection{\label{sec:RevExpt} Review of experimentally observed features}
In this section, we review the experimental features of surfaces subjected to IBSS with singly charged (usually noble gas) ions with kinetic energies ranging from 0.1 to 10 keV. Other singly charged ions have been used in a few cases but are known to produce similar surface morphologies and topographies as the noble gas ions, except for some distortions which are believed to be localised and abnormal [4]. Higher energy apparatus are far more expensive such that the operation costs may outweigh the cost-effectiveness of the IBSS technique. Besides they lead to more pronounced sputtering yields and surface erosion, and irradiation-induced damage of the target material which may override any mechanism for self-organised structures of interest. 

Ref. \cite{24} reported the 1 keV, normal incidence, Ar$^+$ sputtering of Si(1 0 0) surfaces with and without incorporation of Mo atom contaminants simultaneously sputtered from the target-fixing clamps. Nano-dot patterned and smooth surfaces were found in the presence and absence of Mo contaminants, respectively. Further, it was demonstrated in Ref. \cite{25, 26} under contaminant-free conditions that normal incidence sputtering of a Si surface with 0.1 to 0.5 keV Ar$^+$ ions smoothens it. 

In Refs \cite{27, 28}, a systematically controlled contaminant-free environment in the low energy, 200 eV to 1keV, Ar$^+$ ion sputtering of Si surfaces for a wide range of incidence angle $\theta$ was reported. Nanostructures were not found for $\theta$ less than a critical angle $\theta_c\approx 48^o$, nano-ripples with wave vector parallel to the projection of the ion beam direction onto the surface plane, was found for $\theta_c<\theta<80^o$, while nano-ripples with perpendicular wave vector was found for $\theta=85$. $\theta_c$ seems to differ with the target material, for instance larger$\theta_c$ has been found in Ref. \cite{29} for low energy Xe$^+$ sputtering of Si targets. The precise dependence of $\theta_c$ on the target material and sputtering parameters is yet to be ascertained.

Patterning dynamics is known to increase with $\theta$, hence, ripples formed near $\theta_c$ may require longer irradiations times, to be as pronounced as those for $\theta_c>\theta_c$ which may be more prominent for shorter sputtering times. It should be noted that in all the experiments the sputtered material was not removed from the vacuum chamber but mostly redeposited on the target and re-incorporated into the target material.

\subsection{\label{sec:RevTheor} Review of existing discrete models}
In the discrete model of Hartmann et. al. \cite{10, 11, 12} the sputtering process is simulated on a square lattice (surface) of area $L^2$, where L is the lateral size, with periodic boundary conditions. An ion is projected from a random position in a plane parallel to the surface plane, travels along a straight path with an inclination angle of $\theta^o$  to the surface normal and azimuth $\phi$. On reaching the surface, the ion penetrates the target substrate through a depth $d$ along the original trajectory, but now within the target, distributing its energy within an ellipsoid centered on its stopping position d. As a consequence of the energy redistribution, a surface atom at position {\bf r} is eroded, using standard Monte Carlo simulation techniques, with a metropolis algorithm and probability proportional to the energy $E({\bf r})$, received at {\bf r} and given \cite{30} by
\begin{eqnarray}
E({\bf r})=\frac{\varepsilon}{\sqrt{8\pi^3}\sigma\mu^2}exp\left(-\frac{d^2_{\parallel}}{2\sigma^2}-\frac{d^2_{\perp}}{2\mu^2}\right)
\end{eqnarray}

where $\varepsilon$ is the energy of the ion, $d_{}$ is the distance of the surface atom from the stopping point of the ion, and $d_{}$ is the distance of the atom perpendicular to the ion trajectory. $\sigma$ and $\mu$ are the widths of the ellipsoid parallel and perpendicular to the ion trajectory, respectively. This model simulates material redistribution on the surface as a result of the ion impact, and the non-equilibrium thermal interactions of the target, according to a nearest-neighbour solid-on-solid model of surface diffusion in Monte Carlo simulations of surface growth \cite{31}. 

Other researchers have refined the energy distribution $E({\bf r})$, as given in Eq. (1) by the Sigmund model \cite{32, 33}, for a better agreement with MD simulations and experiments. However, these modifications only give very approximate results of the sputter yield at grazing angles. As a matter of fact, the modified Sigmund theory \cite{33} does not produce zero sputter yields for $\theta=90^o$.

Yang et. al. performed kinetic MC (kMC) simulations of ion beam surface sputtering, using Green’s functions obtained from MD simulations, to study the mechanisms of sputter-erosion and surface curvature-dependent sputtering \cite{9}. These Green's functions, which are called Crater functions, accounts for both sputtering and ion-induced re-organisation of surface particles and are used to determine the response of a surface to bombardment with a broad ion beam \cite{13}. These crater functions differ from the linear response functions studied much earlier, in the pioneering stages of the development of a suitable theory of sputter erosion, by Eklund et. al. who concluded that linear response theory is not suitable for the description of the phenomenon of sputter-erosion \cite{2, 34}. The model of Yang et. al. is a hybrid MD-kMC model which added a kMC Arrhenius model \cite{2}, for the simulation of surface diffusion, to existing MD model \cite{35, 36}. 

Nietiadi et. al. performed MC and MD simulations of the sputtering of spherical objects, such as clusters, nanoparticles, or aerosol particles, with energetic particles; taking as a prototypical example, the sputtering of a-Si clusters with 20 keV Ar ions \cite{8}. They employed the MC code of TRI3DST with collisional algorithms that were taken from the sputter version TRIM.SP of the TRIM (TRansport of Ions in Matter) code, with recent modifications \cite{37}. Among other classical motion, the propagation of incident ions and their generated recoils in an amorphous medium was simulated as a sequence of binary collisions in a repulsive screened Coulomb potential with the Kr-C parametrisation \cite{38}, and the interaction of Si atoms of the cluster was modelled with the Stillinger-Weber potential \cite{39} which fits to the Ziegler-Biersack-Littmark potential \cite{40} for small interaction distances.

\subsection{\label{sec:Probs} Some open questions and issues with the existing models}
The issue of contamination, arising from secondary sputtering effects, has rarely being addressed by existing discrete models and remains an open question which was a part of the motivation for this new model, and which this model was aimed at addressing by making it semi-empirical. Although, contamination is an unwanted anomaly in IBSS, it is now been regarded as a new tool to control the properties of the nanostructures as well as for developing new nanostructures. The yield parameter of this model provides an avenue for studying the influence of these contaminants, however, in this first application, we put the contaminant problem aside to first consider the basics. A brief idea of how the model is applicable to the contaminant problem is provided in the suggestion for further work.

The issue of the derivation and parametrisation of an adequate equation of motion for an appropriate continuum modeling of the surface height evolution is also still an open question which this model has the potential of solving via simulated surface height evolution data analysis which can provide detailed information about the height evolution as a function of the relevant sputtering parameters, and can therefore provide knowledge of the applicable discrete and continuum equations of motion.
The kinetic Monte Carlo models have a number of parameters for which realistic or practical interpretations are yet unclear, and seem to be so because they are generalised theories which require extensive simulations for fine-tuning the parameters.  This new model bypasses some of these parameters by being based mostly on the empirical sputtering yield.
More importantly this new model has a direct relation to experimental parameters like type of ion, ion energy, angle of incidence, and target material, through the sputtering yield.

\section{ \label{sec:Model} Relevant Parameters and Construction of the Model}
The parameters of any suitable model are the type of ion, ion energy, angle of incidence, type of material or material surface binding energy, ion flux or fluence, and substrate temperature. The first four parameters are accounted for by the sputter yield, while the last two are accounted for in the simulation time (MC steps) and temperature (Boltzmann energy scale $k_B T$), respectively. However, the target (ambient) temperature is a floating variable which, though incorporated in the simulation “temperature”, is inseparably interwoven with the binding energy in the sense that increasing temperatures merely energise the target atoms and reduce the binding energy; this model ignores this effect in this first instance, if found necessary in further studies then it might be incorporated. This model therefore employs a sputter yield parameter $\Upsilon\left(\theta,E\right)$ which encapsulates both the angular and energy dependence. 

Hence this simple model reiterates the dominance of only one mechanism: the modulated erosion. The modulation entails a huge amount of detail but most of these details eventually lead to the same result of the sputter yield; in other words, a relatively low range of sputter yields account for a wide range of other parameters encompassed in the huge details of the processes leading to the erosion of surface particles. 

This model significantly differs from other MC models in its incorporation of a re-deposition algorithm, in the mould of simulation models of surface growth.  In this wise, a holistic viewpoint is taken where the evolving surface is considered as a Canonical ensemble (number of ions is negligible, relative to the number of substrate atoms), and the relaxation time is assumed to be shorter than the ion arrival times (the disturbance is localised in a relatively tiny area around the incidence spot and hence insignificant on the overall relaxation of the material) for approximating the theory with equilibrium statistical physics principles. Hence, in this model, surface diffusion is assumed to be a thermally activated integral part of surface re-organisation, and is not considered separately. The Canonical assumption may become invalid at high fluences (long sputtering times) and small substrate dimensions and thickness, but it nevertheless is a good starting point.

This model comprises of an Ion Beam Surface Erosion Algorithm (IBSEA), a Redeposition of Eroded Material Algorithm (REMA), and a Surface Reorganization Algorithm (SRA). In the IBSEA, a surface site is randomly chosen and a number $\Upsilon\left(\theta,E\right)$ of surface particles in the vicinity of that site are eroded. All eroded particles are same-plane nearest neighbours, except for surface depressions and protrusions, and the erosion process is subjected to periodic boundary conditions. $\Upsilon\left(\theta,E\right)$ is determined from the table of sputter yield values \cite{41} calculated by using the semi-empirical equation of Seah, Clifford, Green and Gilmore for the sputtering yields of monoatomic elemental solids \cite{42, 43}. Nevertheless, any more accurate sputter yield formula can be assumed for $\Upsilon$. If the chosen site is on a depression then, for $\Upsilon>1$, the topmost atoms of nearest neighbor sites are eroded instead of the same-plane nearest neighbours in order to avoid overhangs. On the other hand, if the chosen site is a surface protrusion then the occupied sites below the vacant nearest neighbor sites are eroded, in addition to the chosen site, for $\Upsilon>1$. 
In the REMA we choose a random direction for re-deposition, and a random re-deposition site between a distance of one lattice parameter and $\frac{2E_S}{mg\Upsilon}\approx 0.2{E'}_S/m\Upsilon\times 10^{-19}m$  from the chosen erosion site; with periodic boundary conditions. Where $\Upsilon\left(\theta,E\right)$ is 1.627, for 1keV Ar$^+$ sputtering of Si at $45^o$ incidence, and $m=46.62\times^{-27}$kg, for Si atom. $E_S=\Upsilon\varepsilon \left({E'}_S=\Upsilon\varepsilon/eV\right)$, and $\Upsilon$ reflects that only a small fraction of $\varepsilon$ contributes to the sputtering, the rest merely disturbs the atoms in the vicinity of the impact point through recoils; we set $\Upsilon\sim 2\times 10^{-17}$. The randomness in the range of the ejected particle (re-deposition site) reflects local variations in the projection angle of the ejected particle, as well as in the fraction of the ion kinetic energy ε transferred to the ejected particle. 
That is, even if the surface atom was ejected at the angle of $\pi/4$ radians for maximum range $0.2E_S/m\Upsilon$, its actual maximum range may still be much less, as a result of its site bonding (lattice coordination) or local interactions with its nearest neighbours; both of which are crucial factors that will determine the fraction of $\varepsilon$ transferred to the ejected particle. In line with the standard approach of aiming for higher lattice coordination in Solid-On-Solid (SOS) simulations, to reflect the tendency of atoms to maximize the surface binding energy, we let the ejected atom stick on the re-deposition site or diffuse to its nearest neighbor site, if vacant; whichever gives the highest coordination number.   
In the SRA we assume a surface atom with single lattice coordination, that is, one with no same-plane nearest neighbour, can diffuse to a vacant nearest neighbor site, where it has a higher lattice coordination. In all cases above (REMA and SRA) where there is selection of, or relocation of a surface atom to, another surface site we implement periodic boundary conditions.
With the implementation of re-deposition, the number of particles of the target material is conserved in this model. However, the overall number of particles in the whole system increases due to the incident ions.

\section{\label{sec:Results} Results and Discussion}

\begin{figure}[!htbp]
\includegraphics[angle=0, width=0.8\textwidth]{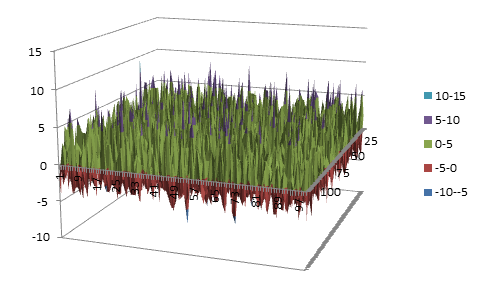}
\caption{Surface profile at $t = 10$. \label{fig:surf10}}
\end{figure}

\begin{figure}[!htbp]
\includegraphics[angle=0, width=0.8\textwidth]{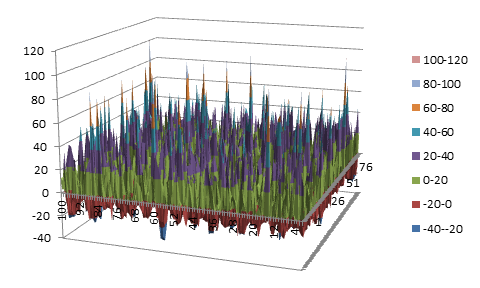}
\caption{Surface profile at $t = 4000$. \label{fig:surf4k}}
\end{figure}

In all results presented here the simulation time is in Monte Carlo steps, which is of the order of ion arrival times, and the distances (e.g. surface height or lateral size) are in lattice spacing, which are of the order of lattice constants. The system size is 100 $\times$ 100 and the simulation was performed with yield parameter of 1keV Ar$^+$ sputtering of Si at 45$^o$ incidence.
Representative figures of the results of the surface topography are as shown in Figs. \ref{fig:surf10} and \ref{fig:surf4k}. Figures \ref{fig:surf10} and \ref{fig:surf4k} are for simulation times of 10 and 4000 Monte Carlo steps, respectively. In both figures (coloured), the legend indicates the height ranges.

Both figures show rough surfaces with no ripple formation, in agreement with the experimental reports reviewed in section \ref{sec:RevExpt}. They also show an increase of the surface depths with sputtering time. Although, it can be observed from these profiles that roughening increases with sputtering time but this is not a sufficient measure. We studied the roughness by calculation of surface roughness.

Surface roughness, or width,  $W$ is a very important index for studying the interplay of competing mechanisms in the surface morphology, such as the destabilising effects of sputter erosion and, in this case, eroded particle re-deposition, and the stabilising effect of reorganisation or diffusion of surface atoms. It is defined as
\begin{equation}W=\sqrt{<\left(h-<h>\right)^2>}\end{equation}

where $h(r)$ is the local surface height at position  $\textbf{r}$, and  $<h>$ is the average surface height at time $t$. $W$ is more conveniently calculated in a simulation by using its simplified form:
\begin{equation}W=\sqrt{<h^2>-<h>^2}\end{equation}

\begin{figure}[!htbp]
\includegraphics[angle=0, width=0.8\textwidth]{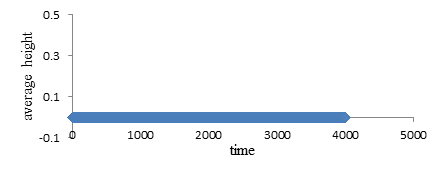}
\caption{Average height of the surface profiles with simulation time.  \label{fig:height}}
\end{figure}

\begin{figure}[!htbp]
\includegraphics[angle=0, width=0.8\textwidth]{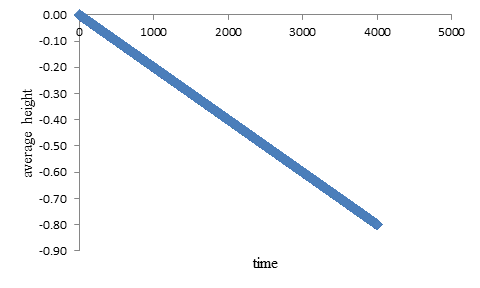}
\caption{Average height of the surface profiles in the absence of redeposition of eroded material. \label{fig:height_noredp}}
\end{figure}

\begin{figure}[!htbp]
\includegraphics[angle=0, width=0.8\textwidth]{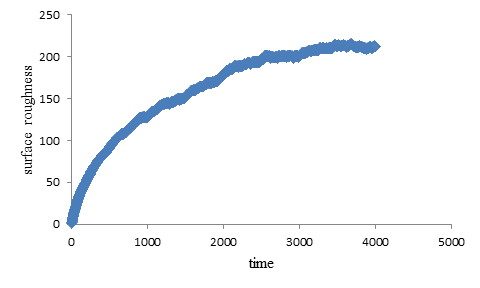}
\caption{Surface roughness as a function of simulation time.  \label{fig:width}}
\end{figure}

We present results of $<h>(t$) and $W(t)$ in Figs. \ref{fig:height} and \ref{fig:width} respectively. Figure \ref{fig:height} shows that the average height of the surface profiles remain constant about a zero value for the entire duration of the simulation. This is to be expected since the number of particles of the target material is constant in this model; that is, as a surface atom is eroded, it is redeposited somewhere else. However, in order to check that this is the only reason, we switched off re-deposition in the simulation and obtained the results shown in Figure \ref{fig:height_noredp}.

Figure \ref{fig:height_noredp} shows the expected result that, when there is no re-deposition of eroded material, the average surface height decreases with simulation time. However, the relatively low values of the average height recorded in Figs. \ref{fig:height} and \ref{fig:height_noredp}, when compared with the higher depth values in Figs. \ref{fig:surf10} and \ref{fig:surf4k}, give a clue as to the importance of the reorganisation process of SRA. Other works \cite{2, 10, 11, 12} based on another robust model \cite{20} exhibits lower values of the average height which may be a reflection of an enhanced reorganization process. 

\begin{figure}[!htbp]
\includegraphics[angle=0, width=0.8\textwidth]{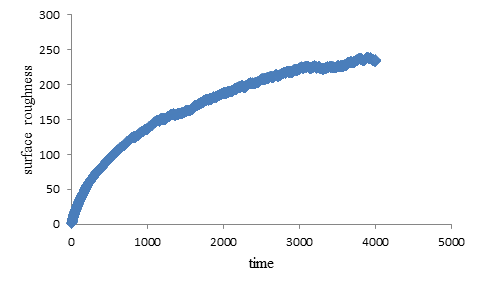}
\caption{Surface roughness without re-deposition of eroded material.  \label{fig:width_noredp}}
\end{figure}

In Figs. \ref{fig:width} and \ref{fig:width_noredp} we present the results of our calculations of the surface roughness with and without re-deposition of eroded material, respectively. We do not observe a significant difference in the roughness for both cases. We believe this to be an indication of the re-deposition process being a ''mirror'' destabilising process of the erosion process, and that the quest for improved lattice coordination by a redeposited particle largely deviates from a filling of depressions generated by the erosion process.

\section{\label{sec:Conclusion} Conclusion and suggestions for further work}

We introduced a new Monte Carlo model that is based on semi-empirical sputter yield in ion-solid energetic collisions, and applied it to 1 keV Argon ion sputtering of Silicon at incidence angle $45^o$. This model circumvents the complexity of the existing statistical, classical and continuum models, most of which are difficult to relate to real experimental parameters, by its reliance on semi-empirical values of the sputter yield. Constrained by this crucial experimental factor, the model addresses the multidimensional nature of other accompanying physical processes stochastically; thus reducing the complexity of their computation. Its simple but realistic algorithms have the potential of being exploited in a hybrid MC-MD (stochastic-classical) simulation algorithm that will reduce the phase volume traversed in MD simulations and allow the hybrid algorithm to simulate the sputtering process up to the timescale of interest in real experiments.

We found that this new model, for Argon ion irradiation of Silicon at $\theta=45^o$, exhibited the experimentally observed features of Silicon surfaces under this condition and allows for a way of studying the effect of the different mechanisms of the surface morphology and the nature of their interplay in the dynamics of surface evolution. Our study of the average surface height revealed that it was constant when eroded particles were redeposited but varied linearly with simulation time, when eroded particles were not redeposited. Our studies also showed that the roughening process was not significantly affected by re-deposition of eroded material.

For further work, yield parameters of other incidence angles could be adopted to study the mechanisms of formation of ripple and nano-dot topographies. As a matter of fact, it might be more convenient to implement the semi-empirical formula of Seah, Clifford, Green and Gilmore directly, if possible, instead of the present form of taking the value of the yield for an ion-target combination as an input from their tables. 

In addition, the independence of the sputter yield of an ion on concurrent sputtering of the target by contaminants should be established to confirm the assumption, here, that the overall yield is a summation of the yield due to the ion and those due to each contaminant that sputters. Then, the erosion stage of this model could be modified to include additional erosion by contaminants for a systematic study of their influence on surface morphology. This could be easily implemented by incorporating an additional empirical yield parameter which captures the yield due to the contaminants but at varying energy, all of which are lower, than the energy of the primary projectile. 

Finally, the model, or its extension, could be used to study the universality class of surface topography evolution by ion bombardment in comparison with other seemingly unrelated phenomena.

\begin{acknowledgments}
The author thankfully acknowledges the research support of the Abdus Salam ICTP in access to current journal articles through their electronic journal delivery system (eJDS). 
\end{acknowledgments}


\begin{thebibliography}{99} 

\bibitem{1}
Taniguchi, N. Tokyo : s.n., 1974, Proc. Intl. Conf. Prod. Eng., Vol. Part II, pp. 18 - 23.

\bibitem{2} 
Oyewande, EO. Modelling and simulation of surface morphology driven by ion bombardment. Institute for Theoretical Physics. Goettingen : University of Goettingen, 2006. pp. 1 - 106, Ph. D. Thesis. Publication URL: https://ediss.uni-goettingen.de/handle/11858/00-1735-0000-0006-B596-C.

\bibitem{3} 
Barabasi, A.-L. and Stanley, H. E. Fractal concepts in surface growth. Cambridge : Cambridge University Press, 1995.

\bibitem{4} 
Self-organized nanopatterning of silicon surfaces by ion beam sputtering. Munoz-Garcia, J, et al., et al. s.l. : Elsevier, 2014, Materials Science and Engineering R, Vol. 86, pp. 1 - 44. And references therein..

\bibitem{5} 
Production of ordered and pure Si nanodots at grazing ion beam sputtering under concurrent substrate rotation. Chowdhury, Debasree, Ghose, Debabrata and Satpati, Biswarup. 2014, Mater. Sci. Eng. B , Vol. 179, pp. 1 - 5. And references therein..

\bibitem{6} 
Subsurface channeling of keV ions between graphene layers: Molecular dynamics simulation. Rosandi, Yudi, Nietiadi, Maureen L and M, Urbassek Herbert. 2015, Physical Review B, Vol. 91, pp. 125441(1-7). and references within..

\bibitem{7} 
Sputter yield of curved surfaces. Urbassek, Herbert M, et al., et al. 2015, Physics Review B, Vol. 91, pp. 165418 (1-9). And references therein..

\bibitem{8} 
Sputtering of Si nanospheres. Nietiadi, Maureen L, et al., et al. 2014, Physical Review B, Vol. 90, pp. 045417 (1-9). And references therein..

\bibitem{9} 
Kinetic Monte Carlo simulation of self-organized pattern formation induced by ion beam sputtering using crater functions. Yang, Zhangcan, Lively, Micheal A and Allain, Jean P. 2015, Physical Review B, Vol. 91, pp. 075427 (1-12). And refernces therein..

\bibitem{10} 
Propagation of ripples in Monte Carlo models of sputter-induced surface morphology. Oyewande, Emmanuel O, Hartmann, Alexander K and Kree, Reiner. 2005, Physical Review B, Vol. 71, pp. 195405 (1-8). And references therein..

\bibitem{11} 
Morphological regions and oblique-incidence dot formation in a model of surface sputtering. Oyewande, Emmanuel O, Kree, Reiner and Hartmann, Alexander K. 2006, Physical Review B, Vol. 73, p. 115434.

\bibitem{12} 
Numerical analysis of quantum dots on off-normal incidence ion sputtered surfaces. Oyewande, Emmanuel O, Kree, Reiner and Hartmann, Alexander K. 2007, Physical Review B, Vol. 75, pp. 155325 (1-8).

\bibitem{13} 
Crater function approach to ion-induced nanoscale pattern formation: Craters for flat surfaces are insufficient. Harrison, Matt P and Bradley, Mark R. 2014, Physical Review B, Vol. 89, pp. 245401 (1 - 10). And references therein..

\bibitem{14} 
Theory of off-normal incidence ion sputtering of surfaces of type AxB1-x and a conformal map method for stochastic continuum models. Oyewande, Emmanuel O and Adeoti, Boluwatife B. 2014, African Review of Physics, Vol. 9, pp. 0024 (1 -7). And references therein..

\bibitem{15} 
Theory of ripple topography induced by ion bombardment. Bradley, Mark R and Harper, James M E. 1988, J. Vac. Sci. Technol. A, Vol. 6, p. 2390.

\bibitem{16} 
Dynamic scaling of ion-sputtered surfaces. Cuerno, Rodolfo and Barabasi, Albert-Laszlo. 1995, Physical Review Letters, Vol. 74, p. 4746.

\bibitem{17} 
Theory of off-normal incidence ion sputtering of surfaces of Type AB and a conformal map method for stochastic continuum models. Oyewande, Oluwole E and Adeoti, Boluwatife B. 2014, African Review of Physics, Vol. 9, pp. 177 - 183.

\bibitem{18} 
Colloquium: Random first order transition theory concepts in biology and physics. Kirkpatrick, T R and Thirumalai, D. 2015, Reviews of Modern Physics, Vol. 87, p. 183.

\bibitem{19} 
Nobel Lecture: The BEH mechanism and its scalar boson. Englert, Francois. 2014, Reviews of Modern Physics, Vol. 86, pp. 845 - 852.

\bibitem{20} 
Long-time effects in a simulation model of sputter erosion. Hartmann, Alexander K, et al., et al. 2002, Physical Review B, Vol. 65, p. 193403.

\bibitem{21} 
Noisy Kuramoto-Sivashinsky equation for an erosion model. Lauritsen, Kent B, Cuerno, Rodolfo and Makse, H A. 1996, Physical Review E, Vol. 54, p. 3577.

\bibitem{22} 
Kinetic Monte Carlo simulations of ion-induced ripple formation: Dependence on flux, temperature, and defect concentration in the linear regime. Chason, Eric, Chan, W L and Bharathi, M S. 2006, Physical Review B, Vol. 74, p. 224103.

\bibitem{23} 
Step formation on the ion-bombarded Ag(100) surface studied by LEED and Monte Carlo simulations. Teichert, Ch, Ammer, Ch and Klaua, M. 2006, physica status solidi(a), Vol. 146, pp. 223 - 242.

\bibitem{24} 
Real-time x-ray studies of MO-seeded Si nanodot formation during ion bombardment. Ozaydin, G, et al., et al. 2005, Appl. Phys. Lett., Vol. 87, p. 163104.

\bibitem{25} 
In situ x-ray studies of native and Mo-seeded surface nanostructuring during ion bombardment of Si(100). Ozaydin-Ince, G and Ludwig Jr, K. 2009, J. Phys. Condens. Matter. , Vol. 21, p. 224008.

\bibitem{26} 
Si(100) surface morphology evolution during normal-incidence sputtering with 100-500 eV Ar ions. Ludwig Jr, F, et al., et al. 2002, Appl. Phys. Lett., Vol. 81, p. 2770.

\bibitem{27} 
Madi, C, et al., et al. 2011, Physical Review Letters, Vol. 107, p. 049902.

\bibitem{28} 
Madi, C, et al., et al. 2008, Physical Review Letters, Vol. 101, p. 246102.

\bibitem{29} 
Macko, S, et al., et al. 2010, Nanotechnology, Vol. 21, p. 085301.

\bibitem{30} 
Sigmund, P. 1973, J. Mater. Sci., Vol. 8, p. 1545.

\bibitem{31} 
Smilauer, P, Wilby, M. R. and Vvedensky, D. D. 1993, Phys. Rev. B, Vol. 47, p. 4119.

\bibitem{32} 
Feix, M, et al., et al. 2005, Phys. Rev. B, Vol. 71, p. 125407.

\bibitem{33} 
A modification to the Sigmund model of ion sputtering. Bradley, Mark R and Hofsaess, Hans. 2014, J. Appl. Phys., Vol. 116, p. 234304.

\bibitem{34} 
Submicron-Scale Surface Roughening Induced by Ion Bombardment. Eklund, E A, et al., et al. 1991, Physical Review Letters, Vol. 67, pp. 1759 - 1762.

\bibitem{35} 
Ion impact crater asymmetry determines surface ripple orientation. Hossain, M Z, et al., et al. 2011, Applied Physics Letters, Vol. 99, p. 151913.

\bibitem{36} 
A multiscale crater function model for ion-induced pattern formation in silicon. Kalyanasundaram, N, Freund, J B and Johnson, H T. 2009, Journal of Physics: Condensed Matter, Vol. 21, p. 224018.

\bibitem{37} 
TRI3DYN - Collisional computer simulation of the dynamic evolution of 3-dimensional nanostructures under ion irradiation. Moeller, Wolfhard. 2014, Nucl. Instrum. Meth. Phys. Res. B, Vol. 322, pp. 23 - 33.

\bibitem{38} 
Calculations of nuclear stopping, ranges, and straggling in the low-energy region. Wilson, W D, Haggmark, L G and Biersack, J P. 1977, Physical Review B, Vol. 15, p. 2458.

\bibitem{39} 
Computer simulation of local order in condensed phases of silicon. Stillinger, Frank H and Weber, Thomas A. 1985, Physical Review B, Vol. 31, p. 5262.

\bibitem{40} 
Ziegler, J F, Biersack, J P and Littmark, U. The Stopping and Range of Ions in Solids. New York : Pergamon, 1985.

\bibitem{41} 
Seah, M P, et al., et al. Surface and Nanoanalysis: Sputter Yield Values. National Physical Laboratory. [Online] August 29, 2014. [Cited: September 29, 2015.] http://www.npl.co.uk/science-technology/surface-and-nanoanalysis/services/sputter-yield-values.

\bibitem{42} 
An accurate semi-empirical equation for sputtering yields I: for argon ions. Seah, M P, et al., et al. 2005, Surf. Interface Anal, Vol. 37, pp. 444 - 458.

\bibitem{43} 
An accurate semi-empirical equation for sputtering yields II: for neon, argon, and xenon ions. Seah, M P. 2005, Nucl. Instrum. Methods Phys. Res. B, Vol. 229, pp. 348 - 358.

\end{thebibliography}

\end{document}